# LEARNING STRUCTURAL COHERENCE VIA GENERATIVE ADVERSARIAL NETWORK FOR SINGLE IMAGE SUPER-RESOLUTION

*Yuanzhuo Li, Yunan Zheng, Jie Chen, Zhenyu Xu, Yiguang Liu*\*

Vision and Image Processing Lab,
School of Computer Science, Sichuan University, China

## ABSTRACT

Among the major remaining challenges for single image super resolution (SISR) is the capacity to recover coherent images with global shapes and local details conforming to human vision system. Recent generative adversarial network (GAN) based SISR methods have yielded overall realistic SR images, however, there are always unpleasant textures accompanied with structural distortions in local regions. To target these issues, we introduce the gradient branch into the generator to preserve structural information by restoring high-resolution gradient maps in SR process. In addition, we utilize a U-net based discriminator to consider both the whole image and the detailed per-pixel authenticity, which could encourage the generator to maintain overall coherence of the reconstructed images. Moreover, we have studied objective functions and LPIPS perceptual loss is added to generate more realistic and natural details. Experimental results show that our proposed method outperforms state-of-the-art perceptual-driven SR methods in perception index (PI), and obtains more geometrically consistent and visually pleasing textures in natural image restoration.

## 1. INTRODUCTION

Single image super-resolution (SISR) aims to recover a high-resolution (HR) image from a single low-resolution (LR) counterpart. It is a highly challenging problem in low-level computer vision task as the ill-posed nature of one-to-many mapping. SISR has been exploited in real-world applications including microscopy imaging and streaming multimedia.

Recently, the field of SISR has been dominated by PSNR-oriented methods [5, 7, 12]. In these studies, they generally adopted L1/L2 loss to increase PSNR. However, since the output of the PSNR-oriented model is the average of the possible SR solutions, their excellent performance in minimizing reconstruction errors always results in blurry images which lack high-frequency information. For this, perceptual-driven methods [15, 22, 27] utilize deep features and adversarial training, achieving a major breakthrough in generating photo-realistic images. While these methods have improved visual quality on a global level significantly, they still suffer from unpleasant details in local textures. One of the sources for the problem lies potentially in the generator network. As the network deepens, some underlying structural information tends to gradually disappear, leading to confusion of high-frequency regions' reconstruction. Besides, the discriminator acting as a classification network in most of state-of-the-art models considers only the whole image or the local per-pixel. Hence it always introduces unstable noise and mottled local structures. Furthermore, several studies focus more on deepening network structures effectively while less on the perceptual loss function. We argue that the performance of widely used VGG based perceptual loss [10] is inadequate due to colored artifacts and inconsistent details.

In this paper, we propose a novel perceptual-driven SR method that includes three aspects to deal with the above-mentioned issues. First, we improve the generator by utilizing gradient branch, which converts the gradient mapping of LR images to that of HR ones as an aid to preserve the image structure. This idea is inspired by [16], we fuse the restored structural information into the SR branch after the upsampling layer to provide explicit structural guidance for the SR process. Our experiments show that this provides the generator with more accuracy in terms of geometry structure. Second, we enhance the discriminator network by introducing decode block into the conventional encode discriminator as in [19]. This U-net structure discriminator simultaneously outputs real and fake decisions that belong to the global image and local per-pixel, which is able to learn a more powerful presentation and leads the generator to pay more attention to the coherence of recovered image globally and locally. We will show that the U-Net discriminator gives more details than normal encode discriminator. Third, we have investigated the loss functions for further improving the visual quality. Specifically, we use LPIPS [28] for perceptual loss instead of VGG based perceptual loss as in the general perceptual-driven methods [15, 16, 22, 23, 27]. We empirically find that it provides more natural textures and more precise details. Experimental results have shown the superiority of our method in both mitigating structural distortions and generating more natural textures compared with state-of-the-art methods.

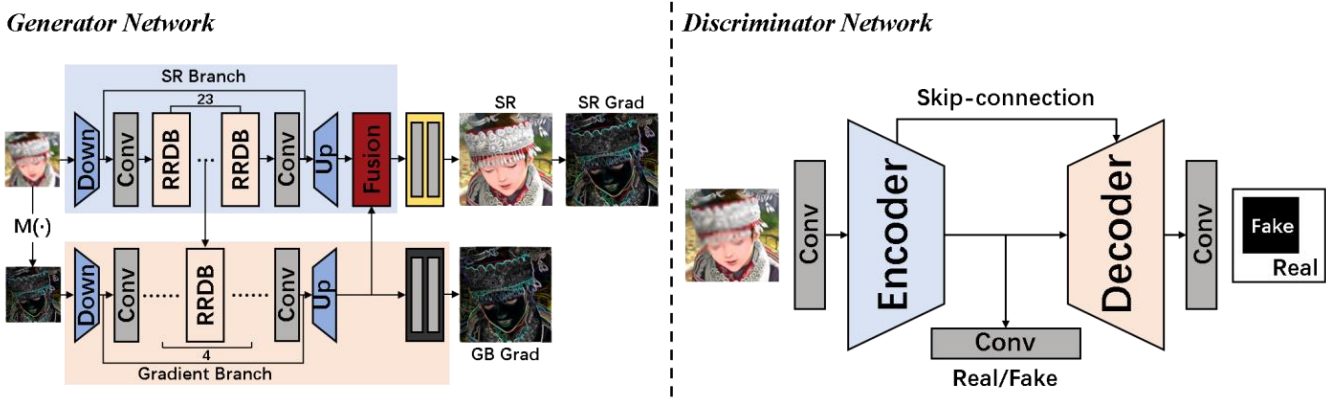

**Fig. 1:** Overall architecture of our Generative Adversarial Network. We adopt RRDB as the high-level architecture of our generator (23 RRDB in SR branch and 4 RRDB in gradient branch). Our U-net based discriminator consists of an encoder part and a decoder part.

## 2. PROPOSED METHOD

### 2.1. Overview

As shown in Fig. 1, our structure guidance generator is composed of two branches, the SR branch and the gradient branch. Besides, our U-net based discriminator has an encoder part and a decoder part. Formally, given input images $I^{LR}$ and a loss function $\mathcal{L}$, our ultimate goal is to generate $I^{SR}$ that are as similar to the corresponding ground-truth images $I^{HR}$ as possible. If generator $G$ parametrized by $\theta_G$, we solve:

$$\hat{\theta}_G = \arg\min \mathbb{E}_{I^{SR}} \mathcal{L}(G(I^{LR}; \theta_G), I^{HR}). \qquad (1)$$

### 2.2. Structure Guidance Generator

We design a structure guidance generator to preserve structures for further improving the image quality, as shown in Fig. 1. For SR branch, the high-level architecture is based on one of the state-of-the-art SR framework RRDB-Net [22]. Benefiting from the dense connections, the SR branch has enough capacity to carry a wealth of structural information. For gradient branch, due to the information of high-frequency is well revealed in the gradient intensity, we use the gradient map of the $I^{LR}$ as input, fitting the spatial distribution of the $I^{HR}$ gradient map through end-to-end learning. The operation to extract gradients is achieved by a fixed convolutional layer $M(\cdot)$ as in [16]. Since the features of the well-designed SR branch contain abundant image prior information, we incorporate 4 (5th, 10th, 15th, 20th) intermediate RRDB features of SR branch to the gradient blocks for further enhancing the efficiency of the gradient reconstruction. Then, we utilize an additional fusion block which contains an RRDB and two convolution layers to fuse the features from the two branches. In general, the performance of the gradient branch is improved by the rich image prior information from the SR branch, and the explicit structural information provided by the gradient branch are concatenated into the SR process to alternately guide the image reconstruction.

### 2.3. U-Net Based Discriminator

Encoder structure discriminator is generally adopted in GAN based SR methods [15, 16, 22, 23, 27], progressively down-sampling the input and capturing the global image context. Inspired by former studies [19, 28], we continuously add a decoder structure to the top of the conventional encoder structure and connect the feature maps between the two modules through skip-connections, see Fig. 1. Our U-net structure discriminator performs real or fake decisions of an input image not just on global but local per-pixel basis, providing comprehensive feedback to the generator for recovering more realistic details. We refer to the U-net discriminator as $D_U$, and discriminator loss $\mathcal{L}_{D_U}$ is computed from both the encoder $D_{Uenc}$ and the decoder $D_{Udec}$:

$$\mathcal{L}_{D_{Uenc}} = -\mathbb{E}_{I^{HR}}[\min(0, -1 + D_{Uenc}(I^{HR}))] \\ - \mathbb{E}_{I^{SR}}[\min(0, -1 - D_{Uenc}(I^{SR}))], \qquad (2)$$

$$\mathcal{L}_{D_{Udec}} = -\mathbb{E}_{I^{HR}}[\min(0, -1 + D_{Udec}(I^{HR}))] \\ - \mathbb{E}_{I^{SR}}[\min(0, -1 - D_{Udec}(I^{SR}))]. \qquad (3)$$

Correspondingly, the generator adversarial loss is as follows:

$$\mathcal{L}_{G_{Adv}} = -\mathbb{E}_{I^{SR}}[D_{Uenc}(I^{SR}) + D_{Udec}(I^{SR})]. \qquad (4)$$

Localization ability matters in U-Net discriminator. Hence, we adopt a consistency regularization technique proposed in [19] to enable it. This technique synthesizes new training data via CutMix augmentation [25] and minimizes the loss $\mathcal{L}_{D_{Ucons}}$, which is shown to be effective for SR problems. In summary, the full discriminator loss is:

$$\mathcal{L}_{D_U} = \mathcal{L}_{D_{Uenc}} + \mathcal{L}_{D_{Udec}} + \mathcal{L}_{D_{Ucons}}. \qquad (5)$$

Therefore, our generator benefits from both the global and the local per-pixel gradients of the images during adversarial training, while in general GAN based SR methods, only one of them takes effect.

## 2.4. Objective functions

The general perceptual-driven SR methods choose VGG based perceptual loss to recover photo-realistic textures [15, 16, 22, 23, 27]. However, VGG network [20] may not be the most appropriate choice for perceptual loss because it is trained on ImageNet [4] classification. Recently, a study found that the AlexNet [13] network is more consistent with the architecture of the human visual cortex [14], while LPIPS [28] reflects human perception preferences more appropriately. Following the work by [9], we use LPIPS trained on the AlexNet network as perceptual loss:

$$\mathcal{L}_{Per}^{SR} = \sum_k \tau(k) \left(\varphi(k)(I^{SR}) - \varphi(k)(I^{HR})\right), \quad (6)$$

here, given an AlexNet network feature extractor $\varphi$, it extracts feature stack from $k$ layers, and $\tau$ transforms deep embedding to a scalar LPIPS score. In addition, we use feature matching loss to alleviate the unpleasant artifacts. $D_{Ul}$ denotes the $l$-th block of the discriminator $D_U$:

$$\mathcal{L}_{fm}^{SR} = \mathbb{E}_{I^{SR}} \|D_{Ul}(I^{SR}) - D_{Ul}(I^{HR})\|_1. \quad (7)$$

Gradient loss has been proposed in [16] to avoid over-smoothing in image restoration. In practice, we use two gradient losses based on the pixelwise for penalizing the two branches of the generator separately. One is $\mathcal{L}_{GM}^{SR_{Pix}}$ that is minimized between the ground-truth gradient map $M(I^{HR})$ and the SR gradient map $M(I^{SR})$:

$$\mathcal{L}_{GM}^{SR_{Pix}} = \mathbb{E}_{I^{SR}} \|M(I^{SR}) - M(I^{HR})\|_1. \quad (8)$$

$$\mathcal{L}_{GB}^{SR_{Pix}} = \mathbb{E}_{I^{SR}} \|I_{GB}^{SR} - M(I^{HR})\|_1, \quad (9)$$

where $\mathcal{L}_{GB}^{SR_{Pix}}$ is the other to optimize the gradient branch, and we design a VGG-Style gradient discriminator $D_{GM}$ additionally to predict relative realness of the gradient map. Thus, we employ RaGAN [11] to achieve this goal:

$$\mathcal{L}_{GM}^{SR_{D_{GM}}} = -\mathbb{E}_{I^{SR}} \left[\log\left(1 - D_{GM}(M(I^{SR}))\right)\right],$$
$$-\mathbb{E}_{I^{HR}} \left[\log\left(D_{GM}(M(I^{HR}))\right)\right]. \quad (10)$$

$$\mathcal{L}_{GM}^{SR_{Adv}} = -\mathbb{E}_{I^{SR}} \left[\log D_{GM}\left(M(G(I^{LR}))\right)\right]. \quad (11)$$

We also add $\mathcal{L}_{Pix}^{SR}$ loss based on the pixel space for accelerating convergence and preventing color permutation:

$$\mathcal{L}_{Pix}^{SR} = \mathbb{E}_{I^{SR}} \|I^{SR} - I^{HR}\|_1. \quad (12)$$

To sum up, we have two discriminators $D_U$ and $D_{GM}$, and optimized by $\mathcal{L}_{D_U}$ and $\mathcal{L}_{GM}^{SR_{D_{GM}}}$ respectively. The total objective functions for our generator and scaling parameters $\lambda$s are as follows:

$$\mathcal{L}_G = \lambda_{Adv}\mathcal{L}_{G_{Adv}} + \lambda_{Per}\mathcal{L}_{Per}^{SR} + \mathcal{L}_{fm}^{SR} + \lambda_{Pix}^{SR}\mathcal{L}_{Pix}^{SR}$$
$$+ \lambda_{Adv}\mathcal{L}_{GM}^{SR_{Adv}} + \lambda_{GM}^{SR}\mathcal{L}_{GM}^{SR_{Pix}} + \lambda_{GB}^{SR}\mathcal{L}_{GB}^{SR_{Pix}}. \quad (13)$$

## 3. EXPERIMENTS

### 3.1 Implementation details

We utilize the DIV2K [1] which consists of 800 images as training dataset. Following former studies [15], we down-sample the HR images by MATLAB bicubic kernel to obtain LR images. In addition, we apply MoA [24] to the training datasets for augmentation. To achieve this, we first up-sample the LR image using nearest kernel to match the HR image size, and then down-sample the image via desubpixel layer [21] at the head of the generator for efficient inference. For testing, we choose five famous benchmark datasets: Set5 [2], Set14 [26], BSD100 [18], Urban100 [8], and General100 [6].

We implement all the experiments in Pytorch 1.7.0 on NVIDIA GTX 1080Ti GPU (11G). For each input mini-batch, we randomly crop 16 128×128 patches from HR images and the corresponding LR image size is 32×32. We employ $\mathcal{L}_{Pix}^{SR}$ to pre-train the generator at first and the learning rate is $2\times 10^{-4}$. The update is performed every $5\times 10^5$ mini-batch cycles with the minimum $1\times 10^{-7}$. We alternately train the generator and discriminator networks about 400k iterations with the total loss functions. The learning rate is set to $1\times 10^{-4}$ and reduced by half at 50k, 100k, 200k, 300k iterations. ADAM optimizer [17] with β1 = 0.9, β2 = 0.999 and ε = $1\times 10^{-8}$ is used for optimization. As for the trade-off parameters of losses, we set $\lambda_{Adv}$, $\lambda_{Per}$, $\lambda_{Pix}^{SR}$, $\lambda_{GM}^{SR}$, and $\lambda_{GB}^{SR}$ to 0.005, 0.001, 0.01, 0.01, and 0.5.

### 3.2 Evaluation and Comparisons

For comprehensively evaluating the perceptual-driven methods, except for PSNR and SSIM, we also choose Perceptual Index (PI) [3] which is more considerable for reflecting human visual preferences. In Table 1, we present quantitative comparisons' results of our methods with state-of-the-art perceptual-driven ones such as SRGAN [15], ESRGAN [22], SFTGAN [23], and USRGAN [27]. We can see that our method achieves the best PI performance in all the testing datasets. However, since our perceptual loss function is not performed on the pixel space but is trained on human judgements, we obtain sharper and more realistic images at the cost of reducing PSNR and SSIM.

In Fig. 2, we visualize some qualitative comparisons of five perceptual-driven SR methods. It is obvious that our method shows more realistic textures without severe structural distortions compared to other methods, as in animal skin and building structures. From the results in USRGAN, it recovers relatively blurry images with the highest PSNR and SSIM, however, our method substantially exceeds USRGAN in PI, indicating that higher PSNR and better visual quality are typically contradictory with each other. The quantitative and qualitative comparisons demonstrate the superiority of our method to recover visually pleasing and photo-realistic SR images compared with state-of-the-art methods.

**Table 1**: Results of quantitative comparison with other methods on five widely used benchmark datasets. Ours method achieves the best PI.

| Dataset | Metrics | Bicubic | SRGAN | SFTGAN | ESRGAN | USRGAN | Ours |
|---|---|---|---|---|---|---|---|
| **Set5** | PI | 7.3699 | 3.5363 | 3.7587 | 3.7525 | 4.1363 | **3.4526** |
| | PSNR/SSIM | 28.420/0.8245 | 29.940/0.8688 | 29.932/0.8665 | 30.454/0.8677 | **30.840/0.8781** | 29.831/0.8539 |
| **Set14** | PI | 7.0268 | 2.9482 | 2.9063 | 2.9261 | 2.9982 | **2.8735** |
| | PSNR/SSIM | 26.100/0.7850 | 26.561/0.7886 | 26.223/0.7854 | 26.276/0.7829 | **27.090/0.8088** | 26.136/0.7807 |
| **BSD100** | PI | 7.0026 | 2.5459 | 2.3774 | 2.4792 | 2.4754 | **2.3627** |
| | PSNR/SSIM | 25.961/0.6675 | 25.498/0.6520 | 25.505/0.6549 | 25.317/0.6506 | **26.020/0.6772** | 25.265/0.6590 |
| **Urban100** | PI | 6.9435 | 3.6952 | 3.6136 | 3.7704 | 3.6208 | **3.5599** |
| | PSNR/SSIM | 23.145/0.9011 | 24.387/0.9376 | 24.013/0.9364 | 24.360/0.9452 | **24.840/0.9497** | 24.144/0.9437 |
| **General100** | PI | 7.9365 | 4.2881 | 4.2878 | 4.3232 | 4.3476 | **3.9919** |
| | PSNR/SSIM | 28.018/0.8282 | 29.358/0.8531 | 29.026/0.8508 | 29.412/0.8546 | **29.870/0.8634** | 29.381/0.8458 |

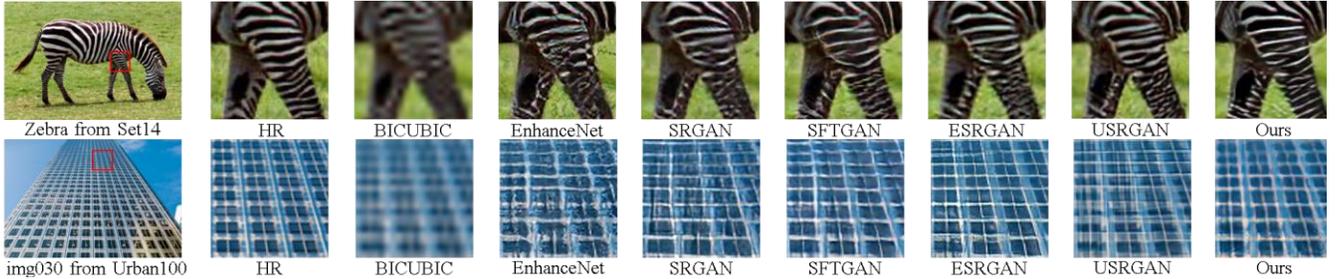

**Fig. 2**: Results of visual comparison on Set14 and Urban100. Our method presents more natural details. Please zoom-in for better view.

**Table 2**: Model configurations and quantitative comparisons. The best performance is achieved by using our full proposed methods.

| Method | Set14 | | | Urban100 | | |
|---|---|---|---|---|---|---|
| | PI | PSNR | SSIM | PI | PSNR | SSIM |
| **Ours w/o GB** | 3.0239 | 26.106 | 0.7731 | 3.6058 | 24.004 | 0.9381 |
| **Ours w/o U-Net** | 2.8771 | 25.828 | 0.7719 | 3.6323 | 24.053 | 0.9404 |
| **Ours w/o LPIPS** | 2.9753 | 26.044 | 0.7774 | 3.6584 | 24.058 | 0.9421 |
| **Ours** | **2.8735** | **26.136** | **0.7807** | **3.5599** | **24.144** | **0.9437** |

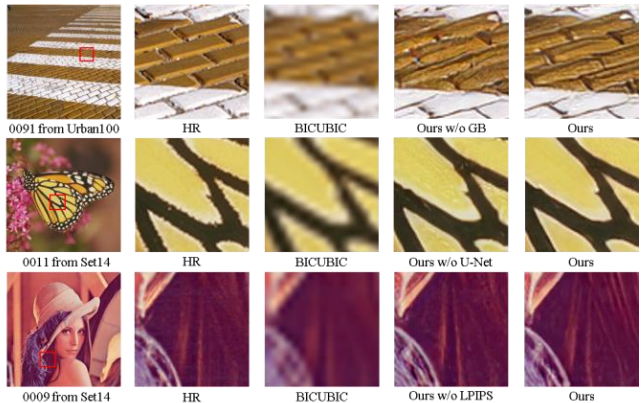

**Fig. 3**: Qualitative comparisons of each component in our methods. Our full proposed method preserves structures better and generates more natural textures.

### 3.3 Ablation Study

In order to validate the necessity of each component in our proposed method, we conduct an ablation study to compare their differences. Detailed model configurations and the quantitative results are depicted in Table 2, corresponding visual results are presented in Fig. 3.

We first remove the gradient branch (GB) in our generator. The quantitative experimental results show that Ours w/o GB reduces the performance of PI significantly. It is also observed in Fig. 3 that the edges of the brick are preserved with the GB better, which demonstrates the effectiveness of the GB in obtaining more information in terms of geometry structures. The second one has the same encode discriminator as the general perceptual-driven methods. We can see that the U-Net discriminator suppresses undesirable noise in the butterfly pattern locally. Besides, the quantitative results suggest that the U-Net can contribute to PI performance. The third is trained without LPIPS but VGG based perceptual loss. As shown in the hair line, Ours w/o LPIPS produces realistic textures while simultaneously generates colored artifacts as well. However, Ours method creates more consistent details and visually pleasing results.

### 4. CONCLUSION

In this paper, we have proposed a new SISR framework to solve the problems of structural distortion and unpleasant textures commonly existing in perceptual-driven SR results. The merits of our work are threefold: 1) The gradient guided generator is shown to exhibit superior performance against structural distortion. 2) A U-Net based discriminator is introduced to provide comprehensive feedback to the generator for recovering more realistic details. 3) An effective perceptual loss based on LPIPS is developed in x4 perceptual SR method to recover more natural textures. These improvements can be combined to balance between the global image and the local texture fidelity in the final output. From the experimental results, it is shown that our proposed method can achieve efficient SISR for natural images.